\documentclass[aps,10pt,superscriptaddress,notitlepage,nofootinbib,floatfix,twocolumn]{revtex4-2}

\usepackage{amsmath}
\usepackage{graphicx}
\usepackage{color}
\usepackage[FIGTOPCAP,raggedright,nooneline]{subfigure}
\usepackage{amsfonts}
\usepackage{dsfont}
\usepackage{multirow}
\newcommand{\abs}[1]{\left| {#1} \right|}

\newcommand{\E}{\mathcal{E}}
\newcommand{\ET}{\abs{\mathcal{E}^{(2)}_0}}
\newcommand{\dint}{\mathrm{d}}
\newcommand{\e}{\mathrm{e}}

\newcommand{\ii}{\mathrm{i}}

\newcommand{\GO}{G_\epsilon^{(0)}}

\begin{document}
\title{Proof of universality in one-dimensional few-body systems including anisotropic interactions}

\author{Lucas Happ}
\email{lucas.happ@uni-ulm.de}
\affiliation{Institut f\"{u}r Quantenphysik and Center for Integrated Quantum Science and Technology ({$\rm IQ$}$^{\rm ST}$),  Universit\"{a}t Ulm, D-89069 Ulm, Germany}
\author{Maxim A. Efremov}
\affiliation{Institut f\"{u}r Quantenphysik and Center for Integrated Quantum Science and Technology ({$\rm IQ$}$^{\rm ST}$),  Universit\"{a}t Ulm, D-89069 Ulm, Germany}
\affiliation{Institute of Quantum Technologies, German Aerospace Center (DLR), D-89069 Ulm, Germany}

\date{\today}

\begin{abstract}
We provide an analytical proof of universality for bound states in one-dimensional systems of two and three particles, valid for short-range interactions with negative or vanishing integral over space.  The proof is performed in the limit of weak pair-interactions and covers both binding energies and wave functions. Moreover, in this limit the results are formally shown to converge to the respective ones found in the case of the zero-range contact interaction.
\end{abstract}

\maketitle

\section{Introduction}
In contrast to purely attractive potentials, which are ubiquitous in quantum physics, interactions whose attractive and repulsive parts cancel each other are only scarcely discussed. Nevertheless, the latter allow for bound states \cite{Simon1976}, and interest in such potentials ramped up within the last years with the ability to realize them in systems of ultracold dipoles \cite{Boettcher2021}. This is supported by recent analysis for these potentials on the formation of few- and many-body bound states in arrays of one-dimensional tubes \cite{Volosniev2013} or in terms of beyond-mean-field contributions in reduced dimensions \cite{Pricoupenko2021}. In two spatial dimensions the scattering properties \cite{Klawunn2010,Rosenkranz2011} have been studied as well as the universality of weakly-bound two-body states \cite{Volosniev2011,Volosniev2011a,Rosenkranz2011} in the experimentally relevant weakly-interacting regime. Here, universality means that the bound states become independent of the details of the interparticle interaction.

In this Letter we study a two- and three-body system of two components, confined to one spatial dimension. We consider only short-range interactions $v(\xi) \equiv v_0 f(\xi)$ with magnitude $v_0$ and shape $f(\xi)$ between distinguishable particles, and none between identical ones. Within these systems we are interested in the universal behavior \cite{Braaten2006} and consider the weakly-interacting limit $v_0 \to 0$, which implies \cite{Simon1976,Gat1993} a weakly-bound two-body ground state. Moreover, we allow for anisotropic features in the interactions which are often present in physical systems.

Within an analytical calculation we prove that in this weakly-interacting limit, interactions of both negative (type I), $\int\dint \xi\, v(\xi) <0$, and vanishing (type II), $\int\dint \xi\,f(\xi) =0$, integral over space lead to the same universal behavior. Our proof is performed for two- and three-body systems alike, by employing the corresponding integral equations in momentum space. The demonstrated universality is not restricted to the binding energies alone, but also includes the corresponding wave functions carrying the full information about the few-body system. In particular, we show that the universal limits are those obtained for a zero-range contact interaction. The energies of three-body bound states $\E_{0,n}$ can then approximately be expressed as
\begin{equation}
\E_{0,n} \simeq \epsilon_n^\star \ET.
\end{equation}
Here, $\E_0^{(2)}$ is the energy of the two-body ground state in the potential $v(\xi)$, and $\epsilon_n^\star$ are the universal energy ratios obtained for the contact interaction. These energy ratios depend only on the heavy-light mass ratio and are presented for several experimentally relevant situations in Ref. \cite{Happ2019}.

This Letter is organized as follows. In section \ref{sec:twobody} we introduce the two types (type I and II) of pair-interactions and present our approach to discuss universality in the two-body domain. We then turn in section \ref{sec:threebody} to the three-body system where we apply a similar approach as for the two-body case in order to prove universality for both types of two-body interactions. Finally, we conclude by summarizing our results and by presenting an outlook in section \ref{sec:conclusion}.

\section{Two interacting particles}\label{sec:twobody}
In this section we first introduce the one-dimensional two-body system and the relevant quantities to describe it. Then we define two different types of interactions whose weakly-bound ground state is discussed in terms of its universality.

\subsection{The two-body system}
In the following, we focus on a two-body system composed of a heavy and a light particle of masses $M$ and $m\leq M$, respectively, all constrained to one spatial dimension. The interaction between both particles is described by a potential
\begin{equation}
v(\xi) \equiv v_0 f(\xi)
\end{equation}
of amplitude $v_0$ and shape $f$. Here, the relative coordinate between the two particles is denoted by $\xi$ in units of the potential range $\xi_0$, while the interaction potential is given in units of the characteristic energy $\hbar^2/(\mu \xi_0^2)$, with the reduced mass $\mu \equiv mM/(m+M)$ and the Planck constant $\hbar$. The pair-interaction is assumed to be real, short-ranged, $\xi^2 v(\xi) \to 0$, as $\abs{\xi} \to \infty$, and to support a bound state.

Moreover, we require that the weakly-interacting limit, $v_0 \to 0$, leads to an even-wave resonance, that is the symmetric part of the two-body ground state wave function becomes dominant compared to the antisymmetric part. We assume that this is the usual case for the ground state of distinguishable particles. There exists of course the counter-example of the odd-wave pseudopotential \cite{Girardeau2004}, however this is a highly nonanalytic model-potential designed to describe the antisymmetric ground state of two non-distinguishable fermions. Finally, we want to emphasize that this requirement is not identical with constraining our analysis to symmetric potentials only, that is we allow for anisotropic features of the potential.

We are only interested in bound states, hence this system is governed by the homogeneous Lippmann-Schwinger equation \cite{Lippmann1950,Sitenko1991}
\begin{equation}\label{eq:LSE_2B_momentum}
\phi^{(2)}(p) =  \frac{v_0}{\E^{(2)} - p^2/2} \int \frac{\dint p'}{2\pi} F(p-p') \phi^{(2)}(p')
\end{equation}
for the two-body wave function $\phi^{(2)}(p)$ in momentum representation. Here
%\begin{equation}
%g_{\E^{(2)}}(p) \equiv \frac{1}{\E^{(2)} - p^2/2}
%\end{equation}
%is the free two-body Green function,
\begin{equation}\label{eq:F_momentum}
F(p) \equiv \int \dint \xi \e^{-\ii p \xi} f(\xi)
\end{equation}
denotes the potential shape $f$ in momentum representation and $\E^{(2)} < 0$ is the total energy of the two-body system in units of $\hbar^2/(\mu \xi_0^2)$.

In the following we consider the weakly-interacting regime, $v_0 \to 0$, which in 1D leads to a weakly-bound ground state \cite{Simon1976,Gat1993} with energy $\E_0^{(2)}$. In order to simplify the analysis we introduce the scaled momentum $P \equiv p/q_0$
%\begin{equation}
%P \equiv \frac{p}{q_0}
%\end{equation}
with
\begin{equation}\label{eq:q0}
q_0 \equiv \sqrt{2\abs{\E_0^{(2)}}}.
\end{equation}
The corresponding integral equation \eqref{eq:LSE_2B_momentum} then reads
\begin{equation}\label{eq:LSE_2B_scaled}
\phi^{(2)}(P) =  -\frac{1}{1+P^2}\frac{v_0}{q_0}  \int \frac{\dint P'}{\pi} F\left[q_0(P-P')\right] \phi^{(2)}(P').
\end{equation}

\subsection{Proof of two-body universality}
Now we prove that in the limit of a vanishing binding energy $\E_0^{(2)}$ of the heavy-light ground state, short-range potentials with both negative, $v_0 F(0) <0$, and vanishing, $F(0)=0$, integral over space yield the same universal solutions for the two-body ground state as for the contact interaction. This universality is shown not only for the binding energy, but also for the corresponding wave function. While this result might not be completely unexpected, the presentation of our approach serves as basis for the subsequent proof of universality in the three-body system (section \ref{sec:threebody}) which is performed in an analogous way.

\subsubsection{Contact interaction}
First, we discuss the case of the zero-range contact interaction of shape $f_\delta(\xi) \equiv \delta(\xi)$, corresponding to $F_\delta(p)=1$. Moreover, for this potential the relation
\begin{equation}\label{eq:q0_contact}
q_0 = -v_0
\end{equation}
remains exact for all values of $v_0$ and $q_0$ \cite{Simon1976}. Hence, in this case the integral equation \eqref{eq:LSE_2B_scaled} takes the form
\begin{equation}\label{eq:LSE_contact_2B}
\phi^{(2)}(P) = \frac{1}{1+P^2} \int \frac{\dint P'}{\pi} \phi^{(2)}(P').
\end{equation}
It is independent of $q_0$ and equivalently independent of $\E_0^{(2)}$, reflecting the scale-invariant property of the delta potential. The solution to this integral equation in the original variable $p$ then takes the form of a Lorentzian
\begin{equation}\label{eq:2BWF_contact}
\phi_\delta^{(2)}(p) = \frac{2 q_0^{3/2}}{q_0^2 + p^2}
\end{equation}
normalized with respect to
\begin{equation}\label{eq:normalization}
\int \frac{\dint p}{2\pi} \left[\phi^{(2)} (p)\right]^2 = 1.
\end{equation}

\subsubsection{Type-I potentials: $v_0 F(0) <0$}
Next, we discuss the potentials with $v_0 F(0)<0$, that is with overall negative integral over space, see Eq.~\eqref{eq:F_momentum}, which we define as type-I potentials. The requirements for the potentials presented in the beginning of section \ref{sec:twobody} still hold. According to Ref. \cite{Simon1976}, we have
\begin{equation} \label{eq:q0_classI}
q_0 = -v_0 F(0) + O(v_0^2)
\end{equation}
as $v_0 \to 0$. Hence, in this limit also $q_0 \to 0$, and the approximation $F\left[q_0(P-P')\right] \simeq F(0)$ can be performed inside the integral in Eq.~\eqref{eq:LSE_2B_scaled}. Thus, by using this approximation together with Eq.~\eqref{eq:q0_classI}, we obtain for Eq. \eqref{eq:LSE_2B_scaled} the same form as Eq.~\eqref{eq:LSE_contact_2B}, that is the same integral equation as for the contact interaction. Consequently, in this limit the normalized wave function $\phi^{(2)}$ converges to the corresponding one $\phi_\delta^{(2)}$, Eq.~\eqref{eq:2BWF_contact}, obtained for the contact interaction.

Effectively, we have used here the fact that in momentum space the potential is more slowly varying around $p'=0$ compared to the wave function, which becomes more localized as $\E^{(2)} \to 0^-$, or $q_0 \to 0$ accordingly. This argument is equivalent to the picture in coordinate representation that in the limit $\E^{(2)} \to 0^-$, the bound state wave function gets broader with respect to the fixed range of the potential.

\subsubsection{Type-II potentials: $F(0) =0$}
Now we analyze potentials for which the integral over space vanishes, that is for which $F(0)=0$, as of Eq.~\eqref{eq:F_momentum}. We denote them as potentials of type II. Here, we additionally require
\begin{equation} \label{eq:convergence}
\frac{\abs{F(p)}^2}{p^2} < \infty \qquad \mathrm{as} \ p \to 0,
\end{equation}
which is always fulfilled for an analytic and smooth potential shape. For these type-II potentials, the linear relation~\eqref{eq:q0_classI} between $q_0$ and $v_0F(0)$ does not hold. Instead, Ref. \cite{Simon1976} derived the quadratic dependence
\begin{equation} \label{eq:q0_classII}
q_0 \simeq  \frac{v_0^2}{\pi} \int \dint p \frac{\abs{F(p)}^2}{p^2}.
\end{equation}

In order to prove that as $q_0 \to 0$ also for the type-II potentials the same solutions as for the contact interactions are retrieved, we iterate Eq.~\eqref{eq:LSE_2B_scaled} once and obtain
\begin{align}\label{eq:LSE_scaled_iterated}
\phi^{(2)}(P) =&~\frac{1}{1+P^2} \frac{v_0^2}{q_0^2} \int \frac{\dint P'}{\pi} \frac{F\left[q_0(P-P')\right]}{1+P'^2} \nonumber \\
&\times \int \frac{\dint P''}{\pi} F\left[q_0(P'-P'')\right] \phi^{(2)}(P'') 
\end{align}
or
\begin{align}\label{eq:LSE_scaled_iterated2}
\phi^{(2)}(P) =&~\frac{1}{1+P^2} \frac{v_0^2}{q_0} \int \frac{\dint P''}{\pi} \phi^{(2)}(P'') \nonumber \\
&\times \int \frac{\dint p'}{\pi} \frac{F\left(q_0 P-p'\right) F\left(p'-q_0P''\right)}{q_0^2 + p'^2},
\end{align}
where we have rescaled the integration variable $P' \equiv p'/q_0$.

In order to perform the limit $q_0 \to 0$ in the integral over $p'$, we have to separately discuss the case $p'=0$. Due to Eq.~\eqref{eq:convergence}, the integrand
\begin{equation}\label{eq:inequality}
\frac{F\left(q_0 P\right) F\left(-q_0P''\right)}{q_0^2} < \infty
\end{equation}
remains finite. For all $p'\neq 0$, the limit $q_0\to 0$ can be performed straightforwardly, hence due to Eq.~\eqref{eq:inequality} we can replace the full integral over $p'$ in Eq.~ \eqref{eq:LSE_scaled_iterated2} by the zero-order Taylor expansion term
\begin{equation} \label{eq:caseii_2B}
\int \frac{\dint p'}{\pi} \frac{F\left(-p'\right) F\left(p'\right)}{p'^2}.
\end{equation}

As a result, we obtain for $q_0 \to 0$ the integral equation
\begin{align}\label{eq:LSE_scaled_iterated3}
\phi^{(2)}(P) =  \frac{4}{1+P^2}\frac{v_0^2}{q_0} \int \frac{\dint p'}{2\pi} \frac{\abs{F(p')}^2}{p'^2} \int \frac{\dint P''}{2\pi} \phi^{(2)}(P'')%\nonumber \\
%& \times \int \frac{\dint P''}{2\pi} \phi^{(2)}(P''),
\end{align}
where we have made use of the fact that $F(-p) = [F(p)]^*$. Application of Eq.~\eqref{eq:q0_classII} in Eq.~\eqref{eq:LSE_scaled_iterated3} then leads to the same integral equation \eqref{eq:LSE_contact_2B} as for the contact interaction. Thus, in the limit $q_0 \to 0$, also the type-II potentials yield solutions $\phi^{(2)}$ which converge to the same limit functions $\phi_\delta ^{(2)}$, Eq.~\eqref{eq:2BWF_contact}, as for the contact interaction.

This concludes the proof of universality in the two-body system for the potentials of type I and II. This universality is equivalent to the statement that in the unitary limit, $\E_0^{(2)}\to0^-$, the corresponding two-body pseudopotential is the delta-potential, even though the delta potential does not feature the property of a vanishing integral over space.

\section{Three interacting particles} \label{sec:threebody}
In this section we first introduce the one-dimensional three-body system which is at the focus of this article. Next, we present a proof of three-body universality that is valid for both type-I and type-II potentials in the weakly-interacting regime.

\subsection{The three-body system}
We now add a third particle to the two-body system, also constrained to 1D and identical to the other heavy particle of mass $M$. We assume the same interaction between the light particle and each heavy one, as introduced in section \ref{sec:twobody}, but no interaction between the two heavy ones.

The homogeneous Lippmann-Schwinger equation~\cite{Lippmann1950,Sitenko1991} governing the bound states in this system can be formulated in complete analogy to the two-body case discussed in section \ref{sec:twobody}
\begin{equation}\label{eq:LSE_repfree}
|\Phi\rangle=  \GO (V_{31} + V_{12})|\Phi\rangle.
\end{equation}
Here however, there are of course two interaction terms, $V_{31}$ and $V_{12}$, corresponding to the interactions of the light particle (particle 1) with each of the two heavy ones (particles 2 and 3). In the center-of-mass-frame of the three-body system, Eq.~\eqref{eq:LSE_repfree} can be cast into the form
\begin{align}\label{eq:LSE_momentum}
& \Phi(P, K)=  \frac{v_0}{q_0}  \GO(P,K) \int \frac{\dint P'}{\pi} F[q_0(P-P')]\nonumber \\
&\ \times \left[\Phi\left(P',K-\frac{P-P'}{2}\right) + \Phi\left(P',K+\frac{P-P'}{2}\right)\right]
\end{align}
where $\Phi(P, K) \equiv \langle P,K|\Phi\rangle$ is the three-body wave function of the two relative motions.

The momenta $P \equiv p / q_0$ and $K \equiv k / q_0$,
%\begin{equation}
%P \equiv p / q_0\ \, \qquad K \equiv k / q_0
%\end{equation}
which are scaled accordingly by $q_0$, Eq.~\eqref{eq:q0}, describe the two relative motions. Indeed, $k$ denotes the relative Jacobi-momentum between the two heavy particles. On the other hand, $p$ is the Jacobi-momentum of the light particle relative to the center of mass of the two heavy ones. We have introduced the free-particle three-body Green function
\begin{equation}
\GO(P,K) = \frac{1}{\epsilon-\alpha_{p} P^{2}-\alpha_{k} K^{2}}
\end{equation}
with the coefficients $\alpha_p \equiv (1+2\alpha)/[2(1+\alpha)]$ and $\alpha_k \equiv 2/(1+\alpha)$ depending only on the mass ratio $\alpha = M/m$. Moreover, the three-body binding energy in units of the energy of the ground state in the heavy-light subsystems is denoted by
\begin{equation}\label{eq:epsilon}
\epsilon \equiv \frac{\E}{\ET}.
\end{equation}
As we only discuss three-body bound states, we restrict the three-body energy and therefore $\epsilon$ to negative values.

%\subsection{Formulation of the problem}
We are interested in the universal \cite{Braaten2006}, that is interaction independent behavior of this three-body system in the weakly-interacting limit $v_0 \to 0$. In particular, we analyze universality of the three-body bound states in terms of the energy spectrum and the corresponding wave functions.

%For the type-I potentials ($v_0F(0)<0$) an analytic proof of universality performed in coordinate-representation has already been presented in Ref. \cite{Happ2019}. The proof states that in the limit $\E_0^{(2)} \to0^-$ both three-body binding energies and corresponding wave functions for any type-I potential converge to the respective results for the contact heavy-light interaction. This proof however cannot be performed in the same way for type-II ($F(0)=0$) potentials. In the following we will present a proof that includes also all type-II potentials.

\subsection{Proof of three-body universality} \label{sec:proof}
For the type-I potentials an analytic proof of universality performed in coordinate-representation has already been presented in Ref. \cite{Happ2019}. This proof however cannot be performed in the same way for type-II potentials. In this subsection we therefore first revisit the original proof \cite{Happ2019} of universality for type-I potentials, however in momentum representation. For this we consider the cases of the contact interaction and any interaction of type I. Next, we extend the proof to type-II potentials.

\subsubsection{Contact interaction}
We start with considering the case of the contact interaction $f_\delta(\xi) = \delta(\xi)$. In this case $F_\delta(p) =1$ and the three-body integral equation \eqref{eq:LSE_momentum} then simplifies with the help of Eq.~\eqref{eq:q0_contact} to the form
\begin{align}\label{eq:LSE_delta}
&\Phi(P, K)=  -\GO(P,K) \int \frac{\dint P'}{\pi} \times \\ \nonumber
& \left[\Phi\left(P',K-\frac{P-P'}{2}\right) + \Phi\left(P',K+\frac{P-P'}{2}\right)\right]. 
\end{align}
Here, $\epsilon$ enters only as a parameter in the Green function $\GO$.

We denote the solutions of Eq.~\eqref{eq:LSE_delta} for the bound-state energy spectrum and the corresponding wave functions by $\epsilon_n^\star$ and $\Phi_n^\star$, respectively. We emphasize that Eq.~\eqref{eq:LSE_delta} is independent of $q_0$ and $\E_0^{(2)}$, therefore the solutions $\epsilon_n^\star$ and $\Phi_n^\star$ are scale-invariant for all values of the two-body binding energy. A more detailed analysis, as well as a table of these energy ratios for a selection of experimentally relevant mass ratios, together with a representation of the full three-body wave functions can be found in Ref. \cite{Happ2019}.

\subsubsection{Type-I potentials: $v_0 F(0) <0$}
Now we discuss the type-I potentials. According to Eq.~\eqref{eq:q0_classI}, the expression $v_0F[q_0(P-P')]/q_0$ present in Eq.~\eqref{eq:LSE_momentum} still converges to $-1$, as $q_0\to 0$. Thus, in this limit we obtain for Eq.~\eqref{eq:LSE_momentum} the same integral equation~\eqref{eq:LSE_delta} as for the contact interaction.

Consequently, as $q_0 \to 0$, the solutions $\epsilon_{0,n}$ and $\Phi_{0,n}$, denoting the three-body energy spectrum and the three-body wave functions for all type-I potentials, converge to the corresponding ones $\epsilon_n^\star$ and $\Phi_n^\star$, obtained for the contact interaction.

\subsubsection{Type-II potentials: $F(0) = 0$}
Next, we present a proof of three-body universality for the type-II potentials. Since in this case $q_0$ is proportional to the second order of $v_0$ and $F$, as summarized by Eq.~\eqref{eq:q0_classII}, we iterate Eq.~\eqref{eq:LSE_momentum} once to the next order in $v_0$ and $F$. In the same spirit as for the two-body system presented in section \ref{sec:twobody}, this then allows us to perform the limit $q_0\to 0$.

Indeed, after iteration Eq.~\eqref{eq:LSE_momentum} takes the form
\begin{equation}\label{eq:LSE_iteration}
\Phi(P,K) = \frac{v_0^2}{q_0} \,  \GO(P,K) \int \frac{\dint P''}{\pi} \left[I_1 + I_2 + I_3 + I_4 \right]
\end{equation}
with
\begin{align}\label{eq:I1234}
&I_1 \equiv \Phi\left(P'',K-\frac{P-P''}{2}\right)\int \frac{\dint P'}{\pi}A_-(P,K,P',P'') \nonumber\\
&I_2 \equiv \int \frac{\dint P'}{\pi}\Phi\left(P'',K-\frac{P+P''}{2}+P'\right) A_-(P,K,P',P'') \nonumber\\ 
&I_3 \equiv \int \frac{\dint P'}{\pi}\Phi\left(P'',K+\frac{P+P''}{2}-P'\right) A_+(P,K,P',P'') \nonumber\\
&I_4 \equiv \Phi\left(P'',K+\frac{P-P''}{2}\right)\int \frac{\dint P'}{\pi}A_+(P,K,P',P''),
\end{align}
and
\begin{align}\label{eq:Kplusminus}
A_\pm(P,K,P',P'') \equiv &~\frac{F[q_0(P-P')]\ F[q_0(P'-P'')]}{q_0} \nonumber\\
&~\times  \GO\left(P',K\pm\frac{P-P'}{2}\right).
\end{align}

We now analyze the expressions $I_{j},\, j=1,2,3,4$. First, we note that in $I_1$ and $I_4$ the argument of $\Phi$ is independent of $P'$. On the other hand, in $I_2$ and $I_3$ the wave function still depends on $P'$ and therefore remains inside the integral. The dependence of $I_{j}$ on $q_0$ can be brought out more clearly by scaling the integration variable $P' \equiv p'/q_0$. The expression
%\begin{equation}
%\frac{A_\pm\left(P,K,\frac{p'}{q_0},P''\right)}{q_0} = \frac{\frac{1}{q_0^2}F(q_0 P-p') F(p'-q_0 P'')}{\epsilon - \alpha_p \frac{p'^2}{q_0^2} - \alpha_k\left[K\pm\left(\frac{P}{2}-\frac{p'}{2q_0}\right)\right]^2}
%\end{equation}
\begin{align}
&\frac{A_\pm\left(P,K,\frac{p'}{q_0},P''\right)}{q_0} =  \nonumber \\
& \qquad \qquad \frac{F(q_0 P-p') F(p'-q_0 P'')}{q_0^2\epsilon - \alpha_p p'^2 - \alpha_k\left[q_0K\pm\frac{1}{4}\left(q_0 P-p'\right)\right]^2}
\end{align}
then appears in each integral of $I_{j}$. For $p'=0$ this expression takes on the value
\begin{equation}\label{eq:pprimeEQzero}
\frac{A_\pm(P,K,0,P'')}{q_0} = \frac{F(q_0 P) F(-q_0 P'')}{q_0^2\left[\epsilon - \alpha_k(K\pm P/2)^2\right]}
\end{equation}
and remains finite also in the limit $q_0 \to 0$, due to Eq.~\eqref{eq:convergence}.

First, we discuss the integrals $I_1$ and $I_4$. Since $A_\pm$ is not singular at $p' =0$, we can replace in $I_1$ and $I_4$ the full integral over $p'$ by the zero-order Taylor expansion term
\begin{equation}\label{eq:pprimeNEQzero2}
\int \frac{\dint P'}{\pi}A_\pm(P,K,P',P'') \to -\int \frac{\dint p'}{\pi} \frac{\abs{F(p')}^2}{p'^2}
\end{equation}
as $q_0 \to 0$. Here we have used the idenitity $\alpha_p + \alpha_k/4 = 1$, and $F(-p) = [F(p)]^*$, as of Eq.~\eqref{eq:F_momentum}. Hence, in the limit $q_0 \to 0$, $I_1$ and $I_4$ are given by
\begin{equation}\label{eq:I1_final}
I_1 \to - \Phi\left(P'',K-\frac{P-P''}{2}\right) \int \frac{\dint p'}{\pi} \frac{\abs{F(p')}^2}{p'^2}
\end{equation}
and
\begin{equation}\label{eq:I4_final}
I_4 \to - \Phi\left(P'',K+\frac{P-P''}{2}\right) \int \frac{\dint p'}{\pi} \frac{\abs{F(p')}^2}{p'^2}.
\end{equation}

Next, we discuss the integrals $I_2$ and $I_3$. Inside the integration over $p'$, there exists the additional factor $\Phi(P'',K\pm(P+P'')/2\mp p'/q_0)$, which eliminates any contribution of the integrand for $\abs{p'}>q_0$, as $q_0 \to 0$. This is because we only discuss normalizable bound states that vanish at infinity, $\Phi(P,K\to \infty) \to 0$. Hence, only the integration in the domain $\abs{p'} \leq q_0$ remains. Since according to Eqs. \eqref{eq:pprimeEQzero} and \eqref{eq:pprimeNEQzero2}, $A_\pm$ is finite therein, we can approximate
\begin{equation}\label{eq:I2approximation}
\abs{I_2} \leq \abs{C} \int\limits_{-q_0}^{q_0} \frac{\dint p'}{\pi}\abs{\Phi\left(P'',K-\frac{P+P''}{2}+\frac{p'}{q_0}\right)}
\end{equation}
with $\abs{C}$ being the maximum value of $\abs{A_\pm}$ inside this integration domain. In order to ensure a finite normalization, the integral $ \iint \dint p\,\dint k \abs{\Phi(p,k)}^2 /(4\pi^2)$
%\begin{equation}
%\iint \frac{\dint p\,\dint k}{4\pi^2} \abs{\Phi(p,k)}^2
%\end{equation}
can have at most a finite contribution from the interval $\abs{p} < q_0$, hence we deduce that the right hand side of Eq.~\eqref{eq:I2approximation}, where $\abs{\Phi}$ enters only linearly, vanishes for $q_0 \to0$. Equivalent arguments can be made also for $I_3$, thus in this limit $I_2 \to 0$ and $I_3 \to 0$.

In total, Eq.~\eqref{eq:LSE_iteration} reduces to
\begin{align}\label{eq:final_inteq}
\Phi&(P,K) = -\GO(P,K) \frac{v_0^2}{q_0} \, \int \frac{\dint p'}{\pi} \frac{\abs{F(p')}^2}{p'^2} \int \frac{\dint P''}{\pi}  \nonumber\\
&\times \left[\Phi\left(P'',K-\frac{P-P''}{2}\right) + \Phi\left(P'',K+\frac{P-P''}{2}\right) \right].
\end{align}
Application of Eq.~\eqref{eq:q0_classII} in this equation then finally leads to the same integral equation \eqref{eq:LSE_delta} as for the contact interaction, which concludes the proof.

As a consequence, in the weakly-interacting limit $v_0\to~0$, also for type-II potentials the solutions $\epsilon_{0,n}$ and $\Phi_{0,n}$ for the three-body energy spectrum and wave functions converge to the corresponding ones $\epsilon_n^\star$ and $\Phi_n^\star$, obtained for the contact interaction.

%\ \newline

\section{Conclusion and Outlook}\label{sec:conclusion}
In the present Letter we have discussed universality of binding energies and wave functions in both the two-body and three-body domain. While in the former the concept of universality for the ground state is often assumed, we have provided here an approach to prove it formally. Application of the same approach to the three-body system has then allowed us to prove universality of the binding energies and wave functions of three-body bound states. In particular, they are shown to converge to the corresponding results for the contact interaction, provided the pair-interactions are tuned to support a weakly-bound two-body ground state. The presented proof of two- and three-body universality is valid for attractive potentials of negative (type I) and vanishing (type II) integral over space alike.

As a result, we can provide approximate expressions for the three-body binding energies 
\begin{equation}
\E_{0,n} \simeq \begin{cases}
-\epsilon_n^\star\, v_0^2\, \left[F(0)\right]^2 & \qquad \text{(type I)\ \ }  \\
-\epsilon_n^\star\, v_0^4\, \left[\displaystyle\int\cfrac{\dint p}{\pi} \cfrac{\abs{F(p)}^2}{p^2}\right]^2 & \qquad \text{(type II)}
\end{cases}
\end{equation}
%\begin{align}
%&\E_{0,n} \simeq -\epsilon_n^\star\, v_0^2\, F(0)^2, \qquad &\text{(type-I)\ \ } \nonumber\\
%&\E_{0,n} \simeq -\epsilon_n^\star\, v_0^4\, \left[\int \frac{\dint p}{\pi} \frac{\abs{F(p)}^2}{p^2}\right]^2 , \qquad &\text{(type-II)}.
%\end{align}
valid in the case of small potential magnitude $v_0$, that is when the pair-interactions are tuned to support a weakly-bound ground state in the heavy-light subsystems.

The universality of energies and wave functions of three-body bound states for finite-range interactions that are tuned to support a weakly-bound \textit{excited} state in the heavy-light subsystems has been demonstrated numerically in Ref. \cite{Happ2021-submitted}. An analytical proof as presented in this work would be desirable and might explain the reported \cite{Happ2021-submitted} differences and similarities compared to the situation of a weakly-bound heavy-light \textit{ground} state.

\acknowledgments
We are very grateful to M.~Zimmermann and W.~P.~Schleich for fruitful discussions. We thank the Center for Integrated Quantum Science and Technology (IQ$^{\rm ST}$) for financial support. The research of the IQ$^{\rm ST}$ is financially supported by the Ministry of Science, Research and Arts Baden-W\"urttemberg.

%\ \newpage 
\appendix

%\section{Appendix I}\label{app:formalism}
%Here might be an appendix.

\bibliography{fewbody}

\end{document}